\documentclass{svproc}
\usepackage{url}

\usepackage{multicol}
\usepackage{footmisc}

\usepackage{amsmath}
\usepackage{amssymb}
\usepackage{graphicx}
\usepackage{subcaption}
\usepackage{color}
\usepackage[colorlinks=true,allcolors=blue]{hyperref}
\usepackage{appendix}
\usepackage{mathtools}
\graphicspath{{figs/}}
\newcommand{\etal}{\emph{et al.}\ }

\newcommand{\ie}{\emph{i.e.}\ }
\newcommand{\eg}{\emph{e.g.}\ }
\newcommand{\resp}{\emph{resp.}\ }

\newcommand{\lambdan}{\lambda^{(n)}}
\newcommand{\mun}{\mu^{(n)}}
\newcommand{\spost}{$s$-post}
\newcommand{\sposts}{$s$-posts}
\newcommand{\cat}{\mathcal{S}}

\DeclareMathSymbol{\Phi}{\mathalpha}{operators}{8} 

\begin{document}
\mainmatter              
\title{Opening up echo chambers via optimal content recommendation}
\titlerunning{Opening up echo chambers}  
%
\author{Antoine Vendeville\inst{1,2}, Anastasios Giovanidis\inst{3}, Effrosyni Papanastasiou\inst{3} and Benjamin Guedj\inst{1,2}}
\authorrunning{Vendeville \etal} 
%
\tocauthor{Antoine Vendeville, Anastasios Giovanidis, Effrosyni Papanastasiou, Benjamin Guedj}
\institute{University College London, Department of Computer Science, Centre for Doctoral Training in Cybersecurity and Centre for Artificial Intelligence, London, UK\\
\email{a.vendeville@ucl.ac.uk},\\ WWW home page:
\texttt{http://antoinevendeville.github.io}\\
\email{b.guedj@ucl.ac.uk}
\and
Inria, Lille - Nord Europe research centre and Inria London\\
\and
Sorbonne University, CNRS, LIP6, F-75005, Paris, France\\
\email{anastasios.giovanidis@lip6.fr}\\
\email{effrosyni.papanastasiou@lip6.fr}}

\maketitle  

\begin{abstract}
Online social platforms have become central in the political debate. In this context, the existence of echo chambers is a problem of primary relevance. These clusters of like-minded individuals tend to reinforce prior beliefs, elicit animosity towards others and aggravate the spread of misinformation. We study this phenomenon on a Twitter dataset related to the 2017 French presidential elections and propose a method to tackle it with content recommendations. We use a quadratic program to find optimal recommendations that maximise the diversity of content users are exposed to, while still accounting for their preferences. Our method relies on a theoretical model that can sufficiently describe how content flows through the platform. We show that the model provides good approximations of empirical measures and demonstrate the effectiveness of the optimisation algorithm at mitigating the echo chamber effect on this dataset, even with limited budget for recommendations.
\end{abstract}

\section{Introduction}
The advent of Online Social Platforms (OSPs) in the last decade has irremediably changed our societies by allowing us to communicate on an unprecedented scale. This has not come without drawbacks, and concerns are rising about potentially nefarious consequences of OSPs \cite{haidt2022}. In particular, the emergence of so-called \emph{echo chambers} have sparked a growing interest in the scientific community. These clusters of like-minded users foster a continuous reinforcement of prior beliefs and strongly reject opposing ideas, thus hindering democratic debate and providing a fertile breeding ground for extremism and conspiracy theories. Occurrences of this phenomenon have been observed in  online discussions surrounding various political and controversial topics \cite{cinelli2021,kirdemir2022,williams2015}, although the amount of users they impact and the extent to which they do so is still up to debate \cite{defranciscimorales2021,dubois2018}. 

Multiple factors have been advanced as explanations for this phenomenon, such as homophily, confirmation bias, negativity or information overload \cite{hills2019,mcpherson2001}. These natural biases are exacerbated by the personalisation algorithms used by the platforms: to sort through the enormous mass of information constantly flowing online, OSPs carefully filter, select and rank only the most relevant content to show to their users, to maximise their engagement. These algorithms tend to hide under the rug anything that support different views, leading to the entrapment of users into their own personalised \emph{filter bubble} \cite{pariser}. This in turn fosters the polarisation of opinions, animosity towards others with different views and the formation of echo chambers \cite{kirdemir2022,santos2021}.



The whole personalisation algorithm used by an OSP is often referred to as a recommendation algorithm. The term may be misleading as these algorithms often mix content the user specifically asked for (by \emph{following} a person or a page) with other items suggested by the platform (`you may be interested in...'). In addition these algorithms may also include filtering or ranking of content. In this paper we use the term \emph{recommendation} specifically to refer to the platform inserting carefully curated content into the newsfeed of users. This mechanism aside we hypothesise that users are presented items as they are posted by their \emph{leaders}---\ie those whom they follow, and do not make any other assumption on the personalisation policy of the platform. 

We introduce a novel method for OSPs to mitigate the echo chamber effect via content recommendation. We start by discussing related works in Section~\ref{literature}. In Section~\ref{problem_statement} we present the dataset \cite{elysee2017_paper} and exhibit the existence of echo chambers therein. To counter it we propose to increase the diversity of content that users are exposed to. We account for the long-term impact of recommendations by extending in Section~\ref{modelling} an existing model \cite{giovanidis2021} that describes how content spreads throughout the network. Section~\ref{optim_section} is then devoted to specifying our mitigation method as a budgeted quadratic program with linear constraints that relies on the model equations. We apply it on the dataset and present the results in Section~\ref{results_section}. We conclude in Section~\ref{conclusion}.

\section{Related literature} \label{literature}
With the advent of OSPs, the impact of algorithmic personalisation on social structures and opinions has attracted an unwavering interest in the scientific community. Many works have studied the consequences of recommending either social connections or content on echo chambers and opinion polarisation, with a variety of metrics proposed to measure these effects. 

Rewiring connections based on similarity, be it of opinion or common acquaintances, tend to foster the emergence of echo chambers \cite{cinus2021,arruda2022,santos2021}. Notably, Cinus \etal \cite{cinus2021} propose a framework to study the impact of recommending either social connections or content on any opinion dynamics model. Presenting too much congenial information to users often leads to a feedback loop as both opinions and recommendations get more extreme \cite{rossi2021}. Perra and Rocha \cite{perra2019} find that the presence of triangles in the graph is a key driver of the emergence of echo chambers under various content recommendation policies. However note that under certain conditions, the OSP algorithm may actually reduce polarisation \cite{cinus2021,arruda2022,ramaciottimorales2021,santos2021}. 
 
To prevent the formation of echo chambers, Musco \etal \cite{musco2018} propose a method to dynamically nudge friendship weights while preserving relevance of content shown to the users. As people are not always willing to change their mind and accept cross-cutting ideas, Garimella \etal \cite{garimella2017c} search for friendship recommendations with high chances of being effective. Finally it may be useful to inject some randomness into the recommender system \cite{rossi2021}.

In this paper we quantify the impact of recommendations by the diversity of content users are exposed to. To derive this metric we extend a previous work \cite{giovanidis2021} that describes the flow of content throughout an OSP. Analogously to Matakos \etal \cite{matakos2020} we seek to maximise the diversity of information presented to users, but we also account for the dynamical flow of content through the network.


\section{Problem statement} \label{problem_statement}
First we present the dataset and propose a measure to quantify the diversity of information users are exposed to. This allows us to unveil the existence of echo chambers and we introduce our paradigm for countering the phenomenon.

\subsection{Dataset}
\begin{figure}[t]
	\centering
	\begin{subfigure}{.5\textwidth}
		\includegraphics[width=\textwidth]{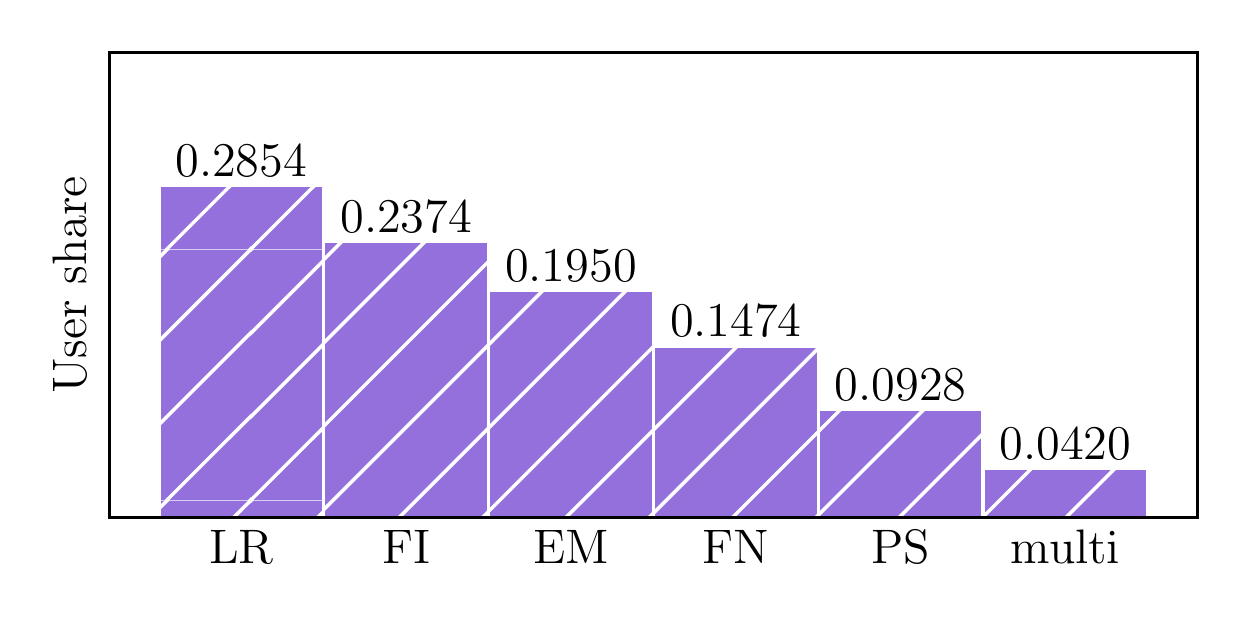}
	\end{subfigure}~
	\begin{subfigure}{.5\textwidth}
		\includegraphics[width=\textwidth]{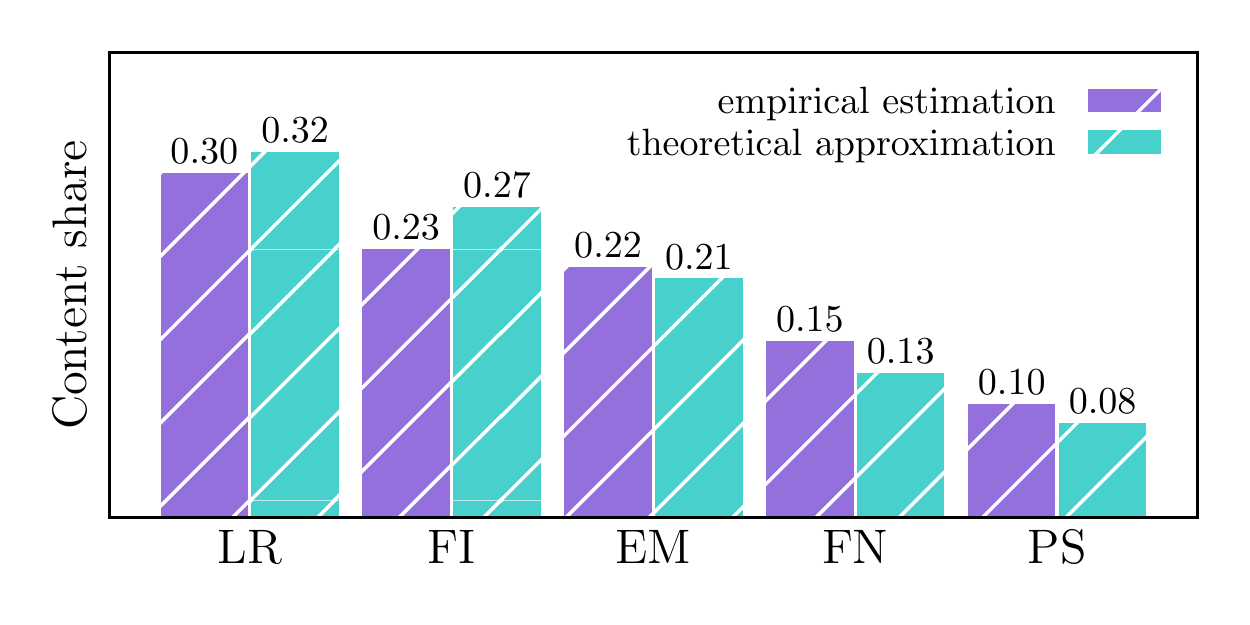}
	\end{subfigure}\\
	\begin{subfigure}{.5\textwidth}
		\includegraphics[width=\textwidth]{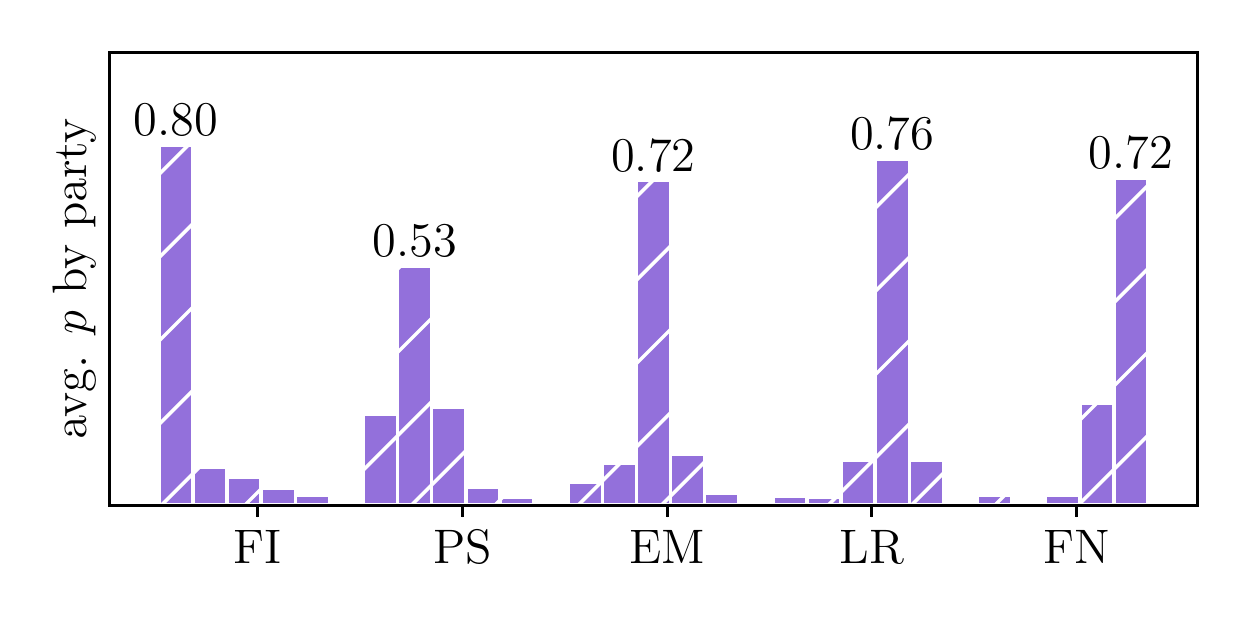}
	\end{subfigure}~
	\begin{subfigure}{.5\textwidth}
		\includegraphics[width=\textwidth]{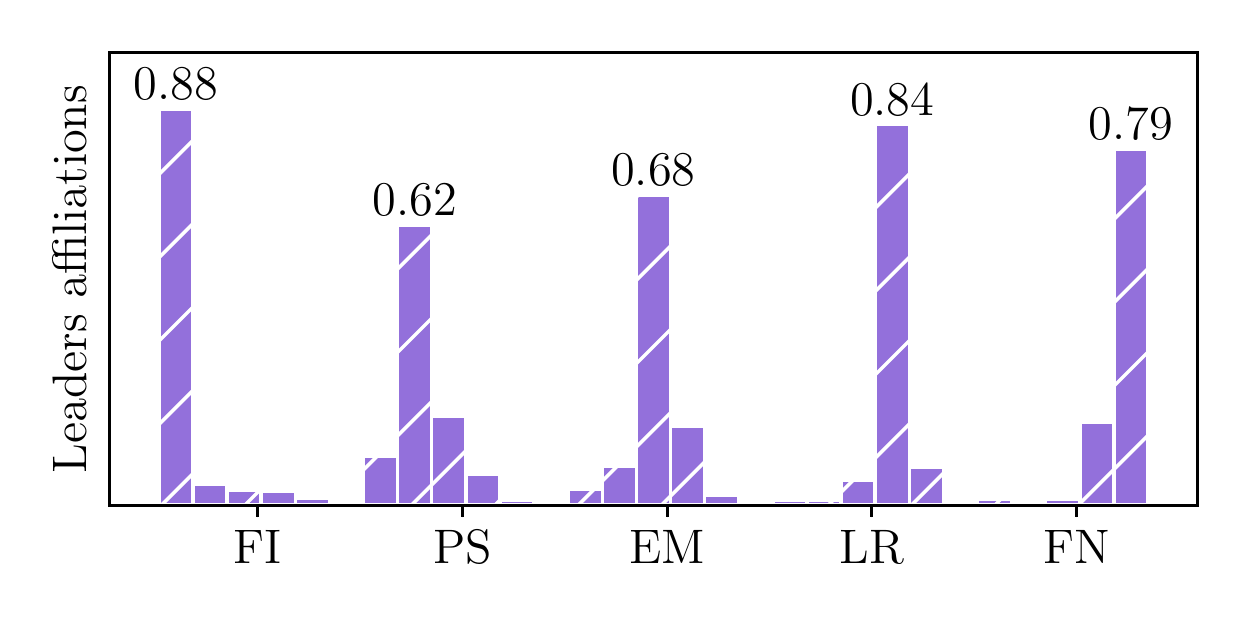}
	\end{subfigure}
	\caption{\textbf{Top left:} user share for each affiliation. \textbf{Top right:} content share for each affiliation. \textbf{Bottom left:} distribution of content on the newsfeeds. \textbf{Bottom right:} distribution of leaders' affiliations by party. Each bar represents a party, in the same order as the x-axis.}
	\label{fig1}
\end{figure}

For the purposes of this work we find it crucial to work with real-world OSP data through which we can quantify the existence of echo chambers. We use the Twitter dataset from \cite{elysee2017_paper} relating to the 2017 French presidential campaign. It includes 2,414,584 tweets and 7,763,931 retweets from 22,853 Twitter profiles ('users' hereafter) discussing the election. Users have been manually annotated by experts with political affiliations describing support for up to two of the main competing parties: 
\begin{description}
	\item[FI] France Insoumise, far-left party (candidate Jean-Luc Mélenchon),
	\item[PS] Parti Socialiste, left-wing party (candidate Benoit Hamon),
	\item[EM] En Marche, centre party (candidate Emmanuel Macron),
	\item[LR] Les Républicains, right-wing party (candidate François Fillon),
	\item[FN] Front National, far-right party (candidate Marine Le Pen).
\end{description} 
For some users the affiliation is unknown and we remove those from our study. Amongst remaining users, a small percentage are affiliated to two different parties (\eg ``PS/EM''). In addition we crawled the followers graph, which was not part of the original dataset. We removed users who did not tweet nor retweeted anything and restricted ourselves to the largest strongly connected component of the followers graph. Finally some profiles were not available anymore and we end up with an anonymised dataset $\mathcal{D}$ that features $N=8,277$ users and $E=975,168$ edges. The share of the considered population for each affiliation is shown in Fig.~\ref{fig1} (top left).

\subsection{Echo chambers} \label{echo_section}
Assume each user is endowed with a \emph{newsfeed} that contains tweets propagated by their \emph{leaders}---\ie other users they follow. Tweets are labelled by the opinion they express, that is support for one of the parties. We define $p^{(n)}_s$ as the average proportion of posts supporting party $s$, thereafter \sposts, on the newsfeed of $n$. The vector $p^{(n)}$ thus describes the distribution of political leanings in the posts user $n$ is exposed to. To quantify the echo chamber effect we look at $echo(n)$, the entry of $p^{(n)}$ corresponding to affiliation of user $n$. The higher this value, the larger share of content presented to $n$ comes from supporters of the same party.

We evaluate $p^{(n)}_s$ empirically for every user $n$ and each party $s$ via the procedure detailed in the Appendix. As seen in Fig.~\ref{fig1} (bottom left), the exposure to congenial content varies between 53\% (PS) and 80\% (FI). The average $echo$ is 0.73, indicating a strong echo chamber effect overall. Notably, about 30\% of users have $echo>0.9$, meaning that almost a third of the online population is barely confronted with opposite ideas. About 5\% of users are even trapped in \emph{perfect} echo chambers, as they have $echo=1$.

These results should not come as too surprising since the user graph exhibits strong homophily, as seen in Fig.~\ref{fig1} (bottom right). Between 62\% (PS) and 88\% (FI) of leaders support the same party, hence congenial information is heavily represented in the newsfeeds.

\subsection{Promoting content diversity}
In order to reduce the echo chamber effect, we seek to increase the diversity of content presented to users on their newsfeeds. In addition we will detail exactly how to do so in practice via personalised recommendations from the platform administrator. To quantify the diversity of the newsfeed of user $n$ we rely on the following metric: 
\begin{equation}
	\Phi_n \overset{\text{def}}{=} \frac{S}{S-1}\sum_{s=1}^S p^{(n)}_s(1-p^{(n)}_s).
\end{equation}
The constant in front of the sum ensures that $\Phi_n$ ranges in $[0,1]$. A value of 0 indicates that the newsfeed of $n$ only contains posts referring to a single party, describing a perfect echo chamber. On the other hand when $\Phi_n=1$ all parties are equally represented on the newsfeed with the same average proportion of $1/S$, meaning a maximally balanced information diet.

Let $\bar\Phi\overset{\text{def}}{=}\sum_n \Phi_n/N$ denote the average diversity of newsfeed over the whole platform. We are looking to solve an optimisation problem of the form
\begin{align} \label{P0} \tag{P0}
	\underset{x,p}{\text{argmax}}\; \bar\Phi
\end{align}
where $x$ are personalised recommendation policies that describe what type of content should be inserted by the platform into the newsfeeds and when. The dependency of $\bar\Phi$ on $x$ is detailed in Section~\ref{optim_section}. Importantly the impact of inserting an item into a newsfeeed is not limited to an immediate change therein, but may also include a broader effect on the whole network as the concerned user can share it to their followers, who may share it further, and so on. The spreading behaviour of the individuals on the platform thus affects the results of the recommendation policies. To model this effect, in the next section we extend an existing model \cite{giovanidis2021} that describes the flow of content throughout the newsfeeds. This allows us to quantify the long-term impact of recommendations on the platform as a whole.

Finally note that we could have chosen to measure diversity via the entropy $-\sum_s p_s^{(n)} \log p_s^{(n)}$ of the newsfeed of $n$. Its use is ubiquitous and its maximum lies in the same place as $\Phi_n$. However the latter has the nice property of being quadratic in $p^{(n)}$ which allows for considerably more efficient optimisation.

\section{Theoretical model} \label{modelling}
The model developed in \cite{giovanidis2021} describes the diffusion of content throughout an OSP. Consider $N$ users who repeatedly create new content (\emph{selfposts}) or spread content created by others (\emph{reposts}). The newsfeed of a user contains posts propagated by their leaders. User $n$ is endowed with two exponential clocks\footnote{Differently distributed times induce similar equilibrium behaviour \cite{giovanidis2021}.}  of respective parameters $\lambdan$ and $\mun$. Whenever the first one rings they create a new selfpost, while the second one prompts them to visit their newsfeed and select an item to repost amongst all content available there. All newsfeeds are of finite size and we make the assumptions that \emph{(i)} the selection of an item to repost when visiting the newsfeed is made uniformly at random, and \emph{(ii)} when a newsfeed is full any new entry will evict an older one chosen uniformly at random\footnote{Other selection and eviction policies induce similar equilibrium behaviour \cite{giovanidis2021}.}. 

The system evolves towards a unique equilibrium characterised by the average state of the newsfeeds, which is solution to a linear system \cite{giovanidis2021}. From there we derive a metric to quantify the overall influence of a user, and compare its values against empirical estimates. The contribution of the present paper related to the original model is twofold. First we introduce opinions by labelling posts with the political party they support. The model is now able to describe opinions dynamics by taking into account not only the topolgoy of the graph but also the activity of its users. We evaluate its performance on the dataset by comparing its predictions with empirical estimates. Second we modify the equilibrium equations to account for the insertion of recommended content into the newsfeeds. This in turn allows us to formulate our method for countering the echo chamber effect as a budgeted quadratic program with linear constraints.


\subsection{Introducing opinions} 
We assume that user $n$ produces selfposts supporting party $s$ at rate $\lambdan_s$, so that $\sum_{s\in\cat}\lambdan_s=\lambdan$. In other words a proportion $\lambdan_s/\lambdan$ of all selfposts from $n$ that refer to party $s$. Accounting for preferences when reposting introduces theoretical difficulties so we leave the mathematical analysis of that case for future research\footnote{Certain contexts might actually benefit from this simplification. For example mentions and replies on Twitter may sometimes represent support and sometimes hostility towards the initial content \cite{williams2015}.}. Let $\mathcal{L}^{(n)}$ be the set of all leaders of user $n$. The average proportion $p_s^{(n)}$ of \sposts\ on the newsfeed of $n$ at equilibrium can be computed via Thm.~\ref{thm_solution}.

\begin{theorem}[Balance of opinions on newsfeeds] \label{thm_solution}
For any party $s$ the average proportions $p_s^{(1)}, \ldots, p_s^{(N)}$ of \sposts\ on each newsfeed at equilibrium are solution of the following linear system:
\begin{align}
	p_s^{(n)}\sum_{k\in\mathcal{L}^{(n)}} (\lambda^{(k)}+\mu^{(k)}) &= \sum_{k\in\mathcal{L}^{(n)}} (\lambda_s^{(k)} + \mu^{(k)}p_s^{(k)}),  \quad n=1,\ldots,N \label{p_formula}.
\end{align}
Assuming the user graph is strongly connected and at least one user has $\lambda>0$, the system has a unique solution.
\end{theorem}
This is a direct reformulation of \cite[eq.\ (8)]{giovanidis2021} with posts no more labeled by their creator but by a political opinion. It is a balance equation that equates the input and output rates of \sposts\ on the newsfeed of $n$ induced by the activity of its leaders. On the left-hand side is the rate at which \sposts\ are evicted at random from the newsfeed to be replaced by fresher content. On the right-hand side is the rate at which \sposts\ propagated by the leaders of $n$ enter the newsfeed. It decomposes in the selfposting rates of leaders of $n$ about party $s$, and the reposting rates of content $s$ from leaders of $n$, through their newsfeeds. Note that $p_s^{(n)}$ is the probability to uniformly select a post about opinion s, in the steady state. Solving eq.~(\ref{p_formula}) relies on matrix inversion which is not tractable for large systems, so we use a power iteration algorithm as described in \cite{giovanidis2021}. 


\subsection{Comparison with empirical results} \label{comparison}
The accuracy of the original model when compared with empirical estimates of $p$ was demonstrated in \cite{giovanidis2021} on two real-life datasets. For the present extension we compute theoretical values of $p$ as given by eq.~(\ref{p_formula}) and compare them with empirical evaluations made in Section~\ref{echo_section}. We estimate $\lambda^{(n)}$ (\resp $\mu^{(n)}$) as the total number of tweets (\resp retweets) posted by $n$ divided by the duration covered by the dataset. To label posts we make the simplifying assumption that all tweets created by a user are labelled as the user's affiliation. In practice people obviously may post about other parties, we leave to future research the estimation of more precise labels based on richer features. 

\begin{figure}[!t]
	\centering
	\includegraphics[width=1\textwidth]{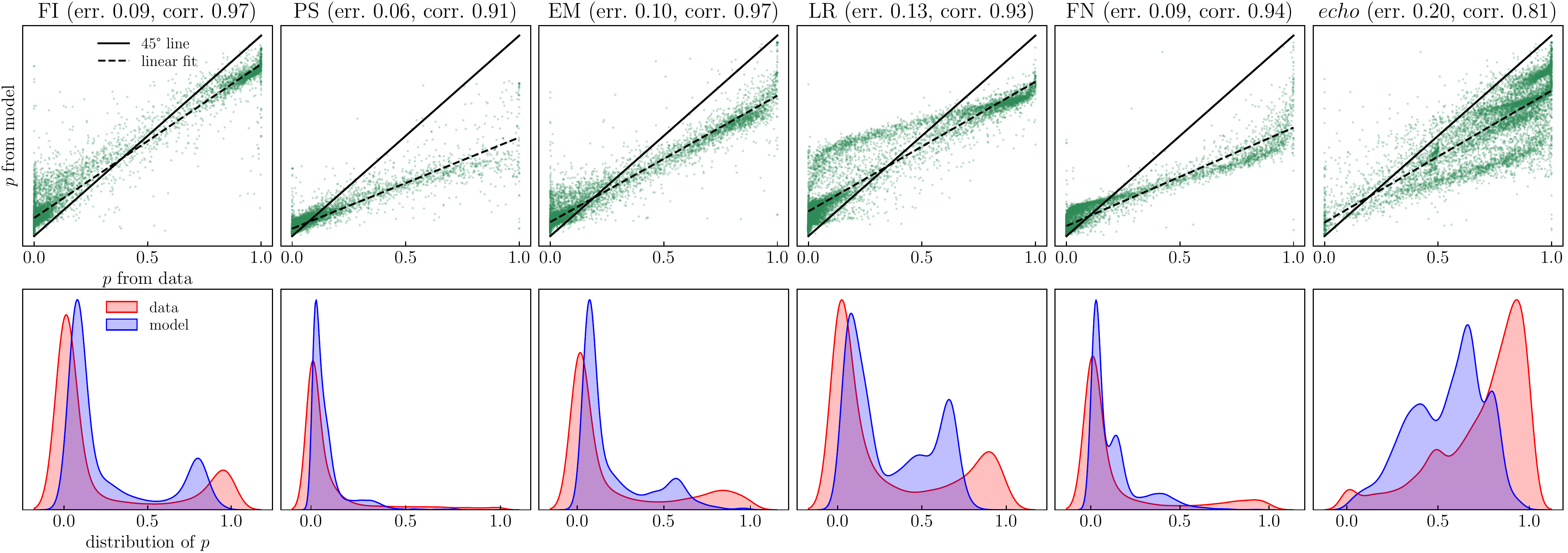}
	\caption{Comparison between $p$ estimated from data and $p$ given by the model. Average errors (`err.') and Pearson correlations (`corr.') between estimations and model are indicated above. \textbf{Top:} scatter plots. \textbf{Bottom:} distributions.}
	\label{modelVSemp}
\end{figure}

We display in Fig.~\ref{modelVSemp} the values of $p_s^{(n)}$ and $echo$ from both the empirical evaluation and the theoretical model. The difference between both is 0.093 on average, and overall tendencies are respected as the Pearson correlation coefficients are close to 1. Moreover the ranking of parties based on their overall share of content, defined for party $s$ as the average of $p_s$ over all users, is the same as for empirical estimations (Fig.~\ref{fig1}, top right). Note that it is also identical to the ranking of parties according to the number of users affiliated to them (Fig.~\ref{fig1}, top left).

However the model is too moderate as it tends to overestimate the small values of $p_s^{(n)}$ and underestimate the high values. The average $echo$ is 0.55 instead of 0.73 evaluated on the data, indicating a lower echo chamber effect. This is not surprising due to the random reposting policy: in the model users repost without distinction for the label of the posts, which is not the case in practice as people are more inclined to interact with congenial content---about 89\% of retweets in the dataset are between supporters of the same party. 

Hence the model captures the echo chamber effect as induced by the graph topology, the differences in posting preferences and the unequal posting and reposting activity. But it does not include the reposting preferences, which as seen in these plots play also an important role in the precise prediction of content in the newsfeeds. Future research shall take this preferential reposting behaviour into account in order to obtain more accurate results.


\section{Maximising content diversity} \label{optim_section}
We are now able to give a precise formulation of problem~(\ref{P0}). Assume that the system can be acted upon by the platform administrator via personalised recommendations, that consist in selecting posts to insert into the newsfeeds of others. Let $x_s^{(n)}$ be the rate at which an \spost\ recommendation is inserted into the newfeed of user $n$ this way. We are looking for the values of $x$ that maximise the overall diversity $\bar\Phi$. For the sake of equity we would like all newsfeeds at equilibrium to contain on average the same proportion $B<1$ of recommended content. Users with very active leaders and thus fast-changing newsfeeds should get new recommendations more often than those with quieter leaders. Formally we require for any user $n$:
\begin{equation} \label{budget_constr}
	\sum_sx_s^{(n)} = \frac{B}{1-B} \sum_{k\in\mathcal{L}^{(n)}} \left(\lambda^{(k)}+\mu^{(k)}\right).
\end{equation}
This ensures that a proportion $B$ of all content arriving on the newsfeed of $n$ is a recommendation. The recommender system can be seen as if each user $n$ has an artificial leader controlled by the platform, who creates \sposts\ at rates $x_s^{(n)}$ and these immediately appear on the newsfeed of user $n$, as recommendations. Hence, the steady-state of the newsfeeds is now given for all $n$ and $s$ by
\begin{align}
	p_s^{(n)}\left(\sum_sx_s^{(n)} + \sum_{k\in\mathcal{L}^{(n)}} (\lambda^{(k)}+\mu^{(k)})\right) &= x_s^{(n)} + \sum_{k\in\mathcal{L}^{(n)}} \left(\lambda_s^{(k)} + \mu^{(k)} p_s^{(k)}\right). \label{new_balance}
	\intertext{This is straightforward from (\ref{p_formula}). Inserting (\ref{budget_constr}) in (\ref{new_balance}) then yields for all $n$ and $s$,}
	\frac{p_s^{(n)}}{1-B}\sum_{k\in\mathcal{L}^{(n)}} \left(\lambda^{(k)}+\mu^{(k)}\right) &= x_s^{(n)} + \sum_{k\in\mathcal{L}^{(n)}} \left(\lambda_s^{(k)} + \mu^{(k)} p_s^{(k)}\right). \label{p_eq_reco}
\end{align}
Such values $p_s^{(n)}$ exist and are unique, as proved in the Appendix. We are now able to formulate our optimisation problem. 



\begin{theorem}[Diversity maximisation]
The optimal recommendation rates that maximise the average diversity of content on the newsfeeds under budget $B$ can be computed via the following quadratic program with linear constraints.
\begin{align} \label{P}
	\underset{x,p}{\textnormal{argmax}}\quad &\bar\Phi \notag \\
	\quad\quad \textnormal{s.t.}	\quad &\dfrac{p_s^{(n)}}{1-B}\sum_{k\in\mathcal{L}^{(n)}} (\lambda^{(k)}+\mu^{(k)}) = x_s^{(n)} + \sum_{k\in\mathcal{L}^{(n)}} (\lambda_s^{(k)} + \mu^{(k)} p_s^{(k)}), \quad \forall n,s, \notag \\
	&\sum_s x_s^{(n)} = \dfrac{B}{1-B} \sum_{k\in\mathcal{L}^{(n)}} (\lambda^{(k)}+\mu^{(k)}), \quad \forall n, \tag{P} \\
	&x_s^{(n)}, p_s^{(n)} \ge 0 \quad \forall n,s. \notag
\end{align}
\end{theorem}
With $N=8,277$ nodes and five categories, problem (\ref{P}) has 82,770 variables and 49,662 linear constraints. Note the presence of $p$ in the optimisation variables, due to its values being dependent on the recommendations $x$.


\section{Results} \label{results_section}
\begin{figure}[!t]
	\centering
	\begin{subfigure}{.33\textwidth}
		\includegraphics[width=\textwidth]{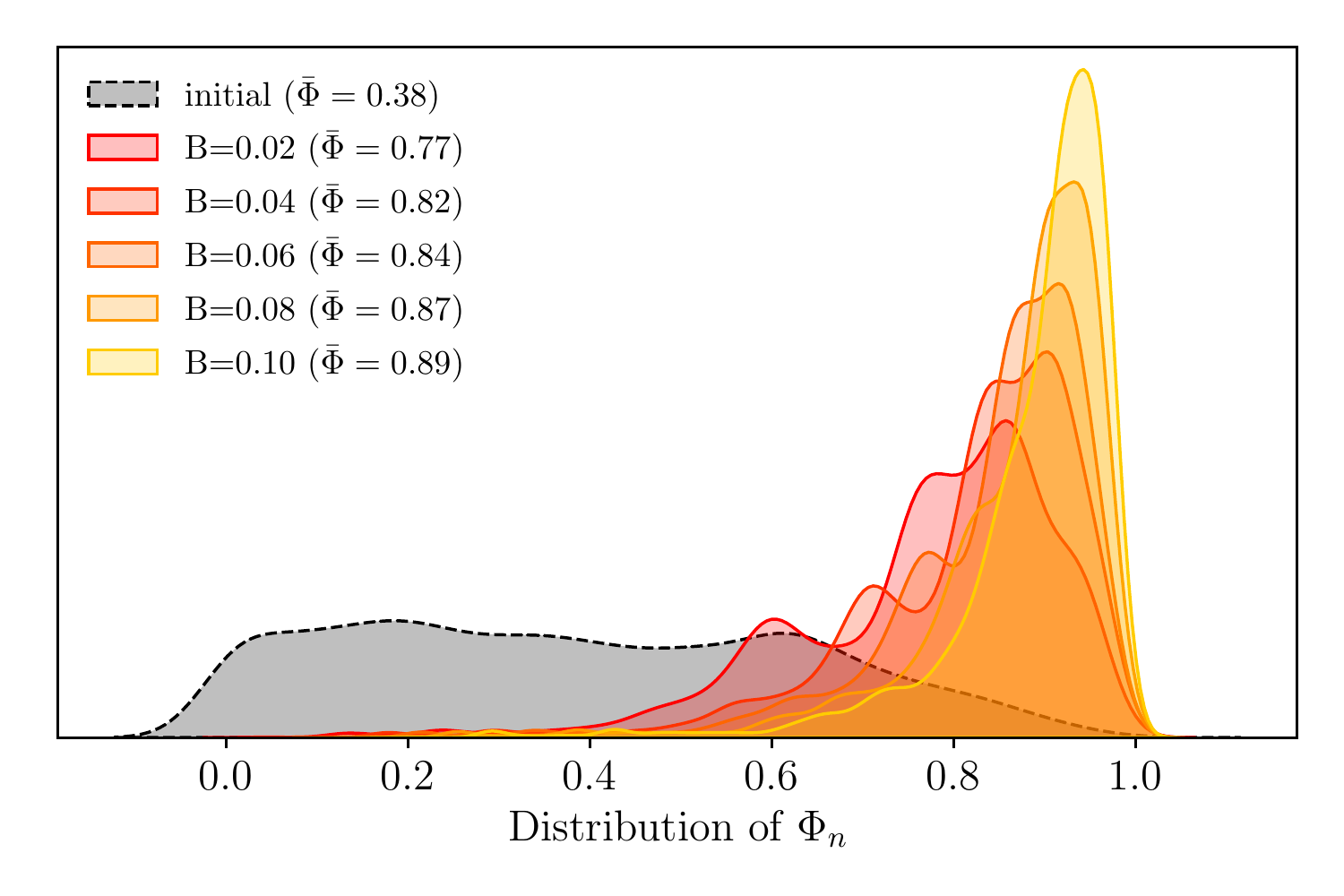}
	\end{subfigure}~
	\begin{subfigure}{.33\textwidth}
		\includegraphics[width=\textwidth]{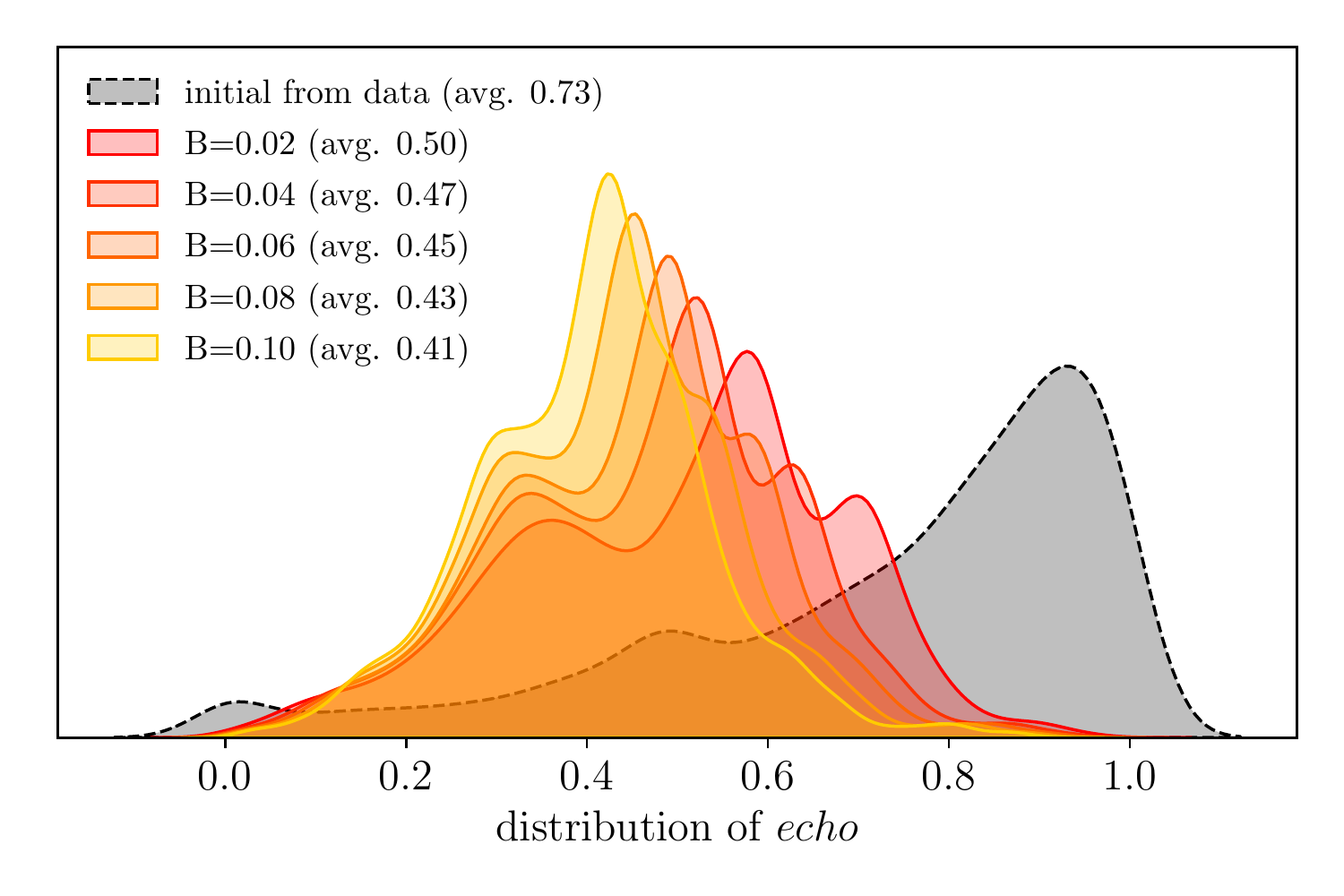}
	\end{subfigure}~
	\begin{subfigure}{.33\textwidth}
		\includegraphics[width=\textwidth]{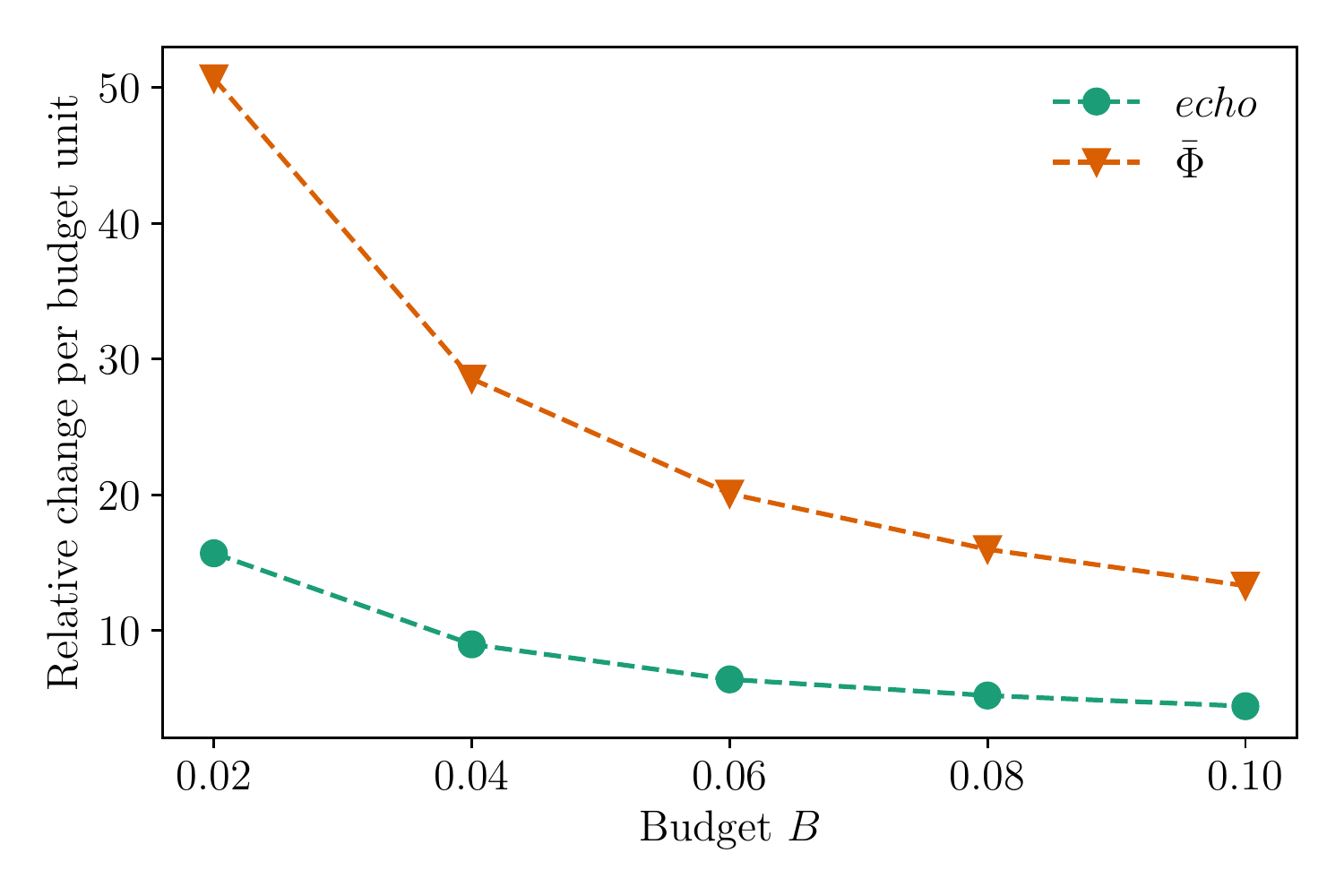}
	\end{subfigure}\\
	\begin{subfigure}{.33\textwidth}
		\includegraphics[width=\textwidth]{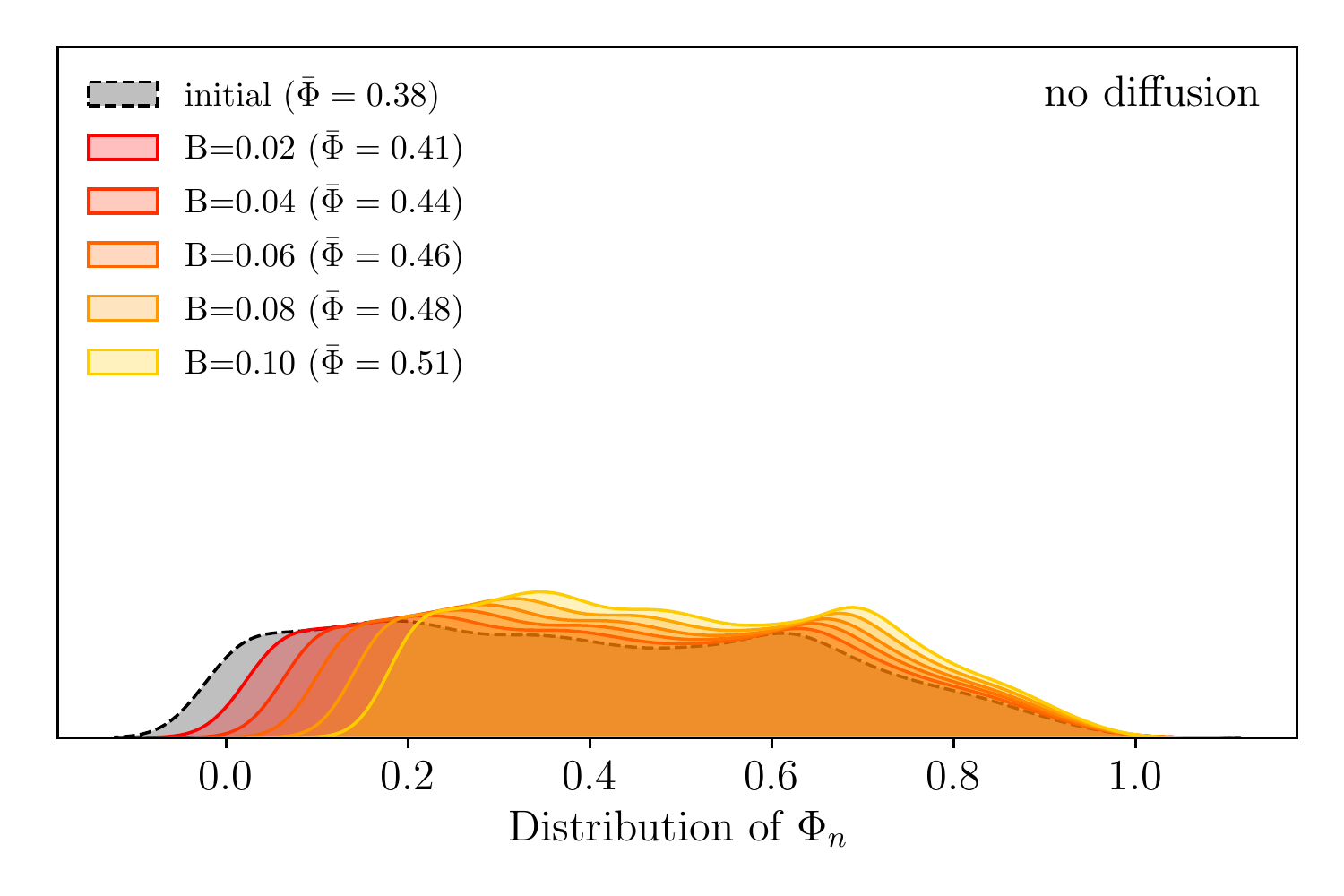}
	\end{subfigure}~
	\begin{subfigure}{.33\textwidth}
		\includegraphics[width=\textwidth]{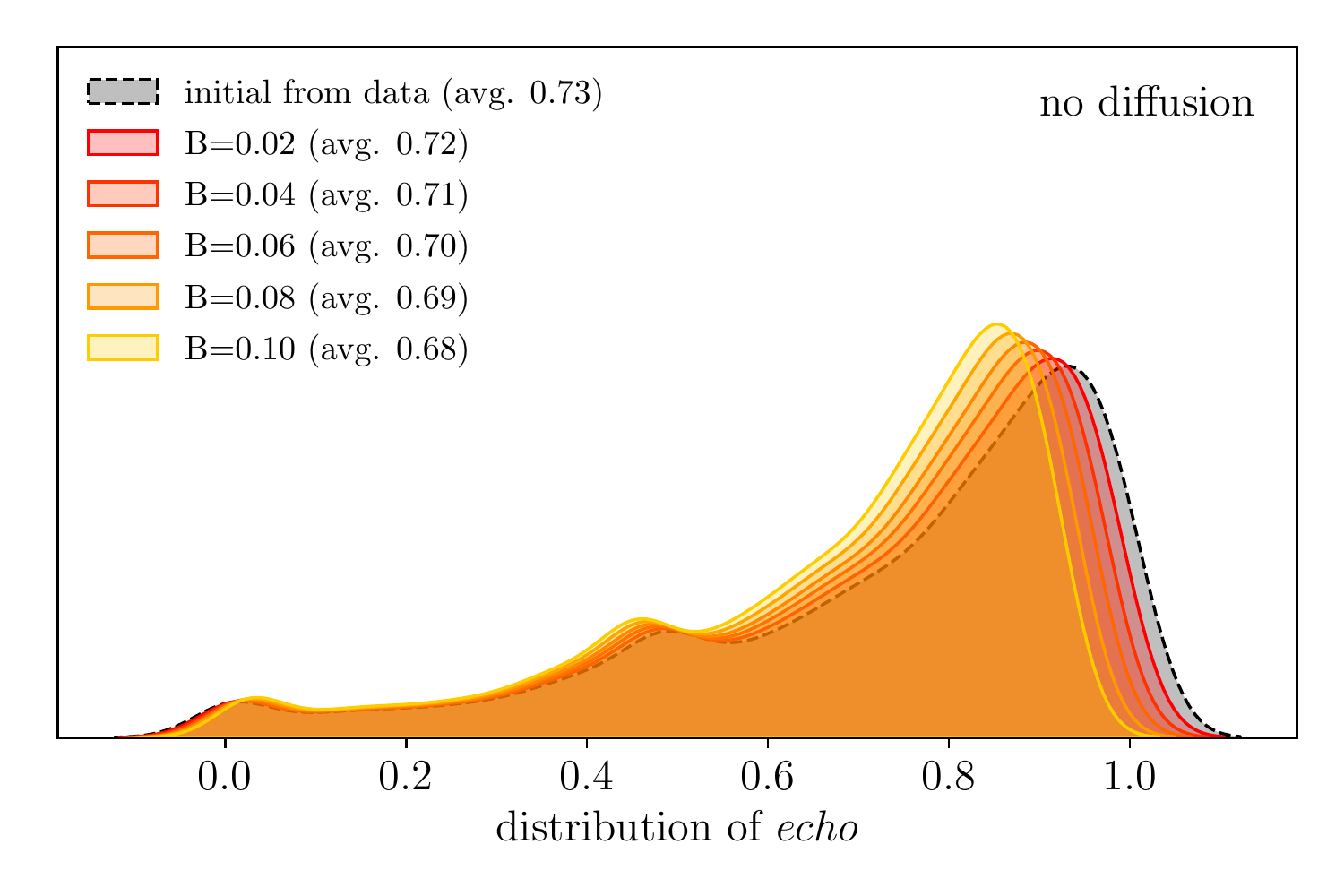}
	\end{subfigure}~
	\begin{subfigure}{.33\textwidth}
		\includegraphics[width=\textwidth]{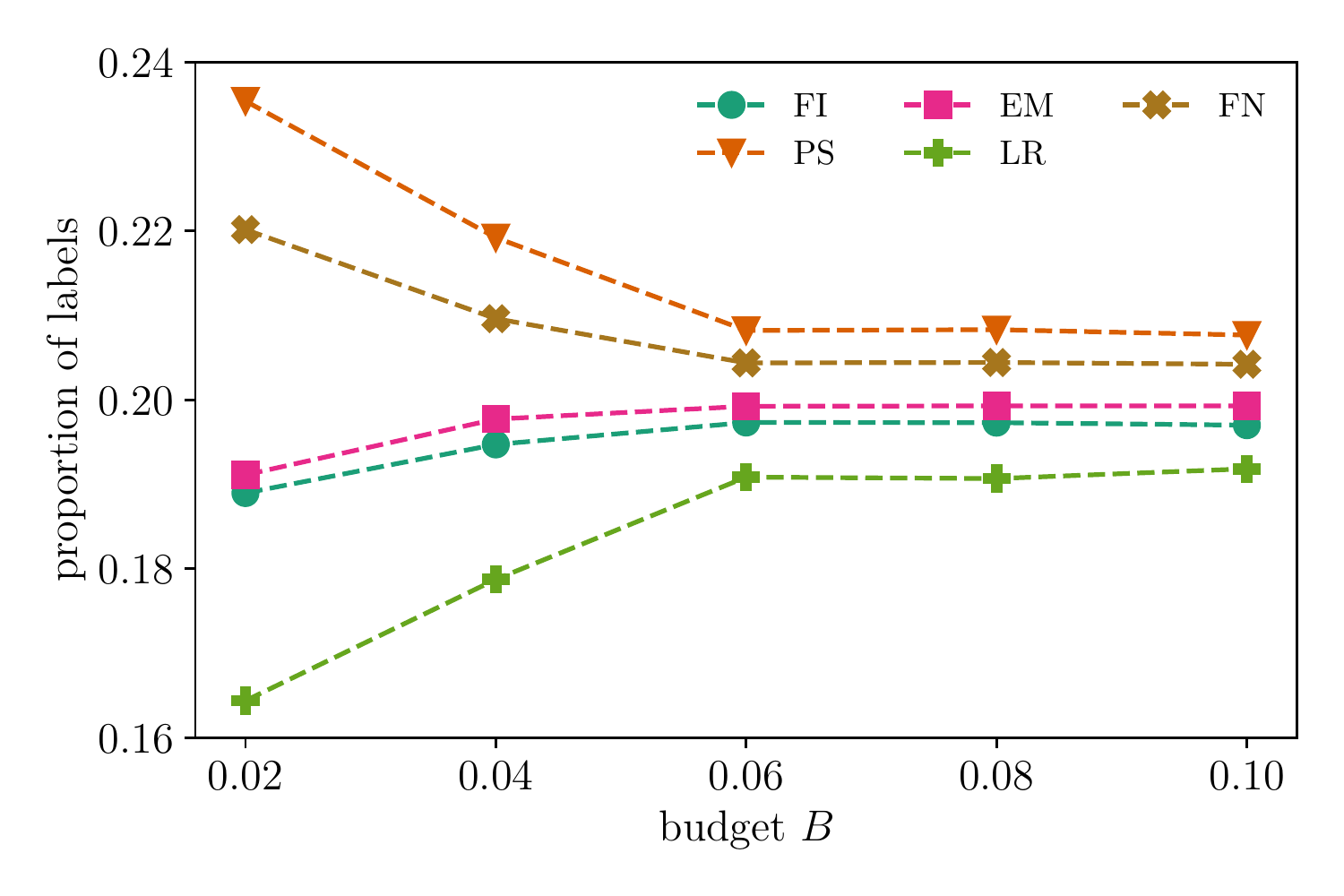}
	\end{subfigure}
	\caption{Optimisation results for various budgets. \textbf{Top:} Distribution of newsfeed diversity (left), echo chambers intensity (middle), and relative impact of each budget unit on $\bar\Phi$ and $echo$ (right). \textbf{Bottom:} results without diffusion of recommended content (left and middle), and distribution of labels in recommended posts (right). Initial distributions correspond to empirical estimates before optimisation.}
	\label{results_plots}
\end{figure}

We solve (\ref{P}) for various values of the budget $B$ and present the results in Fig.~\ref{results_plots}. Experiments are run on a virtual machine with 40 vCPUs and 256GB RAM. For the solution of the optimization problem, we configure a Gurobi solver with the barrier algorithm. The runtime is less than 10 minutes.

Our method is able to drastically improve the diversity of newsfeeds $\Phi_n$ and the echo chamber intensities $echo$ (top left and middle plots). As $B$ increases $\Phi_n$ converges towards 1 and $echo(n)$ towards 0.2 which is $1/S$. An ideally diverse newsfeed has a $p$ vector with all entries equal to $1/S$. The higher the budget the better the results, but small values of $B$ already yield significant improvements. Indeed with just $B=0.02$, meaning users encounter about 2\% of curated recommended content on average, the average newsfeed diversity $\bar\Phi$ is up to 0.77 from 0.38 initially and the average $echo$ is down to 0.50 from 0.73. In the top right plot we show for each value of $B$ the relative change in $\bar\Phi$ and average $echo$ per budget unit. Interestingly, the curves are decreasing, showing that higher budgets have diminishing returns.

The bottom left and middle plot underline how important the diffusion aspect captured by the model is. If we assume that recommended content is never propagated by users, the states of the newsfeeds are given by $p^{(n)}=(1-B)p^{(n)}+B\nu^{(n)}/\Vert\nu^{(n)}\Vert_1$ where $p$ are initial estimates made before optimisation. In that case (bottom left and middle plots), the distributions of $\Phi$ and $echo$ keep their initial shape and their means improve way less.

Finally in the bottom right plot we show the overall share of recommended content that supports each party, that is for each party $s$ we plot $\sum_{n=1}^N \nu_s{(n)}/N$ while the budget varies. These shares, while rather spread at first, converge towards 0.2 as the budget increases. Indeed when $B\rightarrow 1$ the newsfeeds only contain curated recommended content so that maximal diversity is achieved with a balanced representation of all parties. Finally note that the more a party was initially represented in the newsfeeds (fig.~\ref{fig1}, top right), the least they get recommended.

\section{Conclusion} \label{conclusion}
The mitigation of the echo chamber effect through recommendations even with small budgets emphasizes the critical role and responsibility of the social platforms themselves. For the sake of space constraint, we leave for a follow up paper the task of improving the model's accuracy with preferential reposting behaviour. To avoid backfire effects and to preserve their user base, platforms may wish to not feed individuals with too much cross-cutting content. This can be achieved by penalising the change in newsfeeds in the optimisation objective. Moreover because certain parties benefit more from recommendations than others, we may want to enforce equality in the share of recommendations dedicated to each party. Finally we leave for future work the exciting task to consider filtering the content entering a newsfeed or recommending users to one another.




\section*{Acknowledgements}
The authors have no competing interests to declare. This project was funded by the UK EPSRC grant EP/S022503/1 that supports the Centre for Doctoral Training in Cybersecurity delivered by UCL's Departments of Computer Science, Security and Crime Science, and Science, Technology, Engineering and Public Policy. The research of A.G. and E.P. is supported by the French National Agency of Research (ANR) through the FairEngine project under Grant ANR-19-CE25-0011.

\section*{Appendix}

\subsection*{Empirical evaluation of $p$} \label{emp_eval_p}
First we label each tweet by the affiliation(s) of its original creator, or by `?' if the affiliation is unknown. Because the size of the newsfeeds does not matter according to the model \cite{giovanidis2021}, we assume for simplicity that all users have newsfeeds of size 1. We do not know the initial content of each newsfeed and  assume they all contain a post of unknown origin, labelled `?'. As soon as a user tweets or retweets something, the post is inserted into the newsfeeds of its followers, evicting any previous post that was there. Finally to obtain $p_s^{(n)}$ we compute for each user $n$ and label $s$ the proportion of time their newsfeed contained a post labelled $s$. The original affiliation of most tweets is unknown and for each user we disregard periods during which the newsfeed contained a post labelled `?'.  

%

\subsection*{Existence and unicity of $p$ with recommendations}
Let us write equation~(\ref{p_eq_reco}) in matrix form: $p_s = \mathbf{A}p_s +\mathbf{b}.$ As long as the spectral radius $\rho(\mathbf{A})$ of $\mathbf{A}$ is stricly less than 1, the system has a unique solution $p_s = (\mathbf{I}-\mathbf{A})^{-1}\mathbf{b}$. The entries of $\mathbf{A}$ are given by
\begin{equation}
	a_{ij} = (1-B) \frac{\mu^{(j)}}{\sum_{k\in\mathcal{L}^{(i)}} \lambda^{(k)}+\mu^{(k)}} \mathbf{1}_{j\in\mathcal{L}^{(i)}}
\end{equation}
and because $B>0$ it holds that $\sum_j a_{ij}<1$ for any row $i$. But from \cite[Thm.\ 8.1.22]{matrixAnalysis} it we have $\rho(\mathbf{A}) \le \underset{i}{\text{max}}\sum_j a_{ij}$ and thus $p_s$ exists and is unique.
\subsection*{Availability of code}
The code for the optimisation problem is available at\\ \url{https://github.com/AntoineVendeville/Opening-up-echo-chambers}.

%
%
\bibliographystyle{splncs03}
\bibliography{biblio}

\begin{thebibliography}{10}
\providecommand{\url}[1]{\texttt{#1}}
\providecommand{\urlprefix}{URL }

\bibitem{cinelli2021}
Cinelli, M., De~Francisci~Morales, G., Galeazzi, A., Quattrociocchi, W.,
  Starnini, M.: The echo chamber effect on social media. Proc. Natl. Acad. Sci.
   118(9),  e2023301118 (Mar 2021)

\bibitem{cinus2021}
Cinus, F., Minici, M., Monti, C., Bonchi, F.: The effect of people recommenders
  on echo chambers and polarization. arXiv:2112.00626 [physics]  (Dec 2021)

\bibitem{defranciscimorales2021}
De~Francisci~Morales, G., Monti, C., Starnini, M.: No echo in the chambers of
  political interactions on {{Reddit}}. Sci Rep  11(1),  2818 (Dec 2021)

\bibitem{dubois2018}
Dubois, E., Blank, G.: The echo chamber is overstated: the moderating effect of
  political interest and diverse media. Information, Communication \& Society
  21(5),  729--745 (May 2018)

\bibitem{arruda2022}
{Ferraz de Arruda}, H., Maciel~Cardoso, F., {Ferraz de Arruda}, G.,
  R.~Hern{\'a}ndez, A., {da Fontoura Costa}, L., Moreno, Y.: Modelling how
  social network algorithms can influence opinion polarization. Information
  Sciences  588,  265--278 (Apr 2022)

\bibitem{elysee2017_paper}
Fraisier, O., Cabanac, G., Pitarch, Y., Besançon, R., Boughanem, M.:
  {\#Élysée2017fr: The 2017 French Presidential Campaign on Twitter}. In:
  Proceedings of the 12th International AAAI Conference on Web and Social Media
  (2018), dataset available at
  \url{https://zenodo.org/record/5535333\#.Yp974OxBxH0}

\bibitem{garimella2017c}
Garimella, K., De~Francisci~Morales, G., Gionis, A., Mathioudakis, M.: Reducing
  controversy by connecting opposing views. In: Proceedings of the Tenth ACM
  International Conference on Web Search and Data Mining. pp. 81--90. WSDM '17,
  Association for Computing Machinery, New York, NY, USA (Feb 2017)

\bibitem{giovanidis2021}
Giovanidis, A., Baynat, B., Magnien, C., Vendeville, A.: Ranking online social
  users by their influence. IEEE/ACM Transactions on Networking  29(5),
  2198--2214 (2021)

\bibitem{haidt2022}
Haidt, J.: Why the past 10 years of american life have been uniquely stupid.
  The Atlantic  (2022),
  \url{https://www.theatlantic.com/magazine/archive/2022/05/social-media-democracy-trust-babel/629369/},
  accessed on June 6, 2022

\bibitem{hills2019}
Hills, T.T.: The dark side of information proliferation. Perspect Psychol Sci
  14(3),  323--330 (2019)

\bibitem{matrixAnalysis}
Horn, R.A., Johnson, C.R.: Matrix Analysis. Cambridge University Press (1990)

\bibitem{kirdemir2022}
Kirdemir, B., Agarwal, N.: Exploring {Bias} and {Information} {Bubbles} in
  {YouTube}’s {Video} {Recommendation} {Networks}. In: Complex {Networks} \&
  {Their} {Applications} {X}. pp. 166--177 (2022)

\bibitem{matakos2020}
Matakos, A., Tu, S., Gionis, A.: Tell me something my friends do not know:
  Diversity maximization in social networks. Knowl Inf Syst  62(9),  3697--3726
  (Sep 2020)

\bibitem{mcpherson2001}
McPherson, M., Smith-Lovin, L., Cook, J.M.: Birds of a feather: Homophily in
  social networks. Annu Rev Sociol  27(1),  415--444 (2001)

\bibitem{musco2018}
Musco, C., Musco, C., Tsourakakis, C.E.: Minimizing polarization and
  disagreement in social networks. In: Proceedings of the 2018 World Wide Web
  Conference. pp. 369--378. WWW '18, International World Wide Web Conferences
  Steering Committee, Republic and Canton of Geneva, CHE (Apr 2018)

\bibitem{pariser}
Pariser, E.: The Filter Bubble: What the Internet Is Hiding from You. The
  Penguin Group (2011)

\bibitem{perra2019}
Perra, N., Rocha, L.E.C.: Modelling opinion dynamics in the age of algorithmic
  personalisation. Sci Rep  9(1),  7261 (May 2019)

\bibitem{ramaciottimorales2021}
Ramaciotti~Morales, P., Cointet, J.P.: Auditing the effect of social network
  recommendations on polarization in geometrical ideological spaces. In: RecSys
  '21: 15th ACM Conference on Recommender Systems. Amsterdam, Netherlands (Sep
  2021)

\bibitem{rossi2021}
Rossi, W.S., Polderman, J.W., Frasca, P.: The closed loop between opinion
  formation and personalised recommendations. IEEE Transactions on Control of
  Network Systems pp. 1--1 (2021)

\bibitem{santos2021}
Santos, F.P., Lelkes, Y., Levin, S.A.: Link recommendation algorithms and
  dynamics of polarization in online social networks. Proc Natl Acad Sci USA
  118(50),  e2102141118 (Dec 2021)

\bibitem{williams2015}
Williams, H.T.P., McMurray, J.R., Kurz, T., Hugo~Lambert, F.: Network analysis
  reveals open forums and echo chambers in social media discussions of climate
  change. Global Environmental Change  32,  126--138 (May 2015)

\end{thebibliography}

\end{document}